\documentclass[twocolumn,prb,aps,psfig,showpacs,preprintnumbers,superscriptaddress]{revtex4}
\usepackage{graphicx}
\usepackage{epstopdf}
\usepackage{float}
\usepackage{color}
\usepackage{amsmath}

\begin{document}

\title{Volovik Effect and Fermi-Liquid Behavior in the $s$-Wave Superconductor CaPd$_2$As$_2$: $^{75}$As NMR-NQR Measurements} 
\author{Q.-P.~Ding}
\affiliation{Ames Laboratory, U.S. DOE, and Department of Physics and Astronomy, Iowa State University, Ames, Iowa 50011, USA}
\author{P.~Wiecki}
\affiliation{Ames Laboratory, U.S. DOE, and Department of Physics and Astronomy, Iowa State University, Ames, Iowa 50011, USA}
\author{V.~K. Anand}
\affiliation{Ames Laboratory, U.S. DOE, and Department of Physics and Astronomy, Iowa State University, Ames, Iowa 50011, USA}
\affiliation{Helmholtz-Zentrum~Berlin~f\"ur Materialien~und~Energie,~Hahn-Meitner~Platz~1,~D-14109~Berlin,~Germany}
\author{N.~S.~Sangeetha}
\affiliation{Ames Laboratory, U.S. DOE, and Department of Physics and Astronomy, Iowa State University, Ames, Iowa 50011, USA}
\author{Y.~Lee}
\affiliation{Ames Laboratory, U.S. DOE, and Department of Physics and Astronomy, Iowa State University, Ames, Iowa 50011, USA}
\author{D.~C.~Johnston}
\affiliation{Ames Laboratory, U.S. DOE, and Department of Physics and Astronomy, Iowa State University, Ames, Iowa 50011, USA}
\author{Y.~Furukawa}
\affiliation{Ames Laboratory, U.S. DOE, and Department of Physics and Astronomy, Iowa State University, Ames, Iowa 50011, USA}

\date{\today}

\begin{abstract} 
    The electronic and magnetic properties of the collapsed-tetragonal CaPd$_2$As$_2$ superconductor (SC) with a transition temperature of 1.27 K have been investigated by $^{75}$As nuclear magnetic resonance (NMR) and nuclear quadrupole resonance (NQR) measurements.
    The temperature ($T$) dependence of the nuclear spin lattice relaxation rates (1/$T_1$) and the Knight shifts indicate the absence of magnetic correlations in the normal state.        
     In the SC state,  1/$T_1$ measured by $^{75}$As NQR shows a clear Hebel-Slichter (HS) peak just below $T_{\rm c}$ and decreases exponentially at lower $T$, confirming a conventional $s$-wave SC. 
      In addition, the Volovik effect, also known as Doppler shift effect, has been clearly evidenced by the observation of the suppression of the HS peak with applied magnetic field.

\end{abstract}

\maketitle


  Recently considerable experimental and theoretical attention has concentrated on transition-metal pnictides after the discovery of superconductivity (SC) in LaFeAsO$_{1-x}$F$_x$. \cite{Kamihara2008}
    Among the various iron-pnictide superconductors, much attention has been paid to doped $A$Fe$_2$As$_2$ ($A$ = Ca, Ba, Sr, Eu) known as $``$122$"$ compounds. \cite{Johnston2010,Canfield2010,Stewart2011}
     The parent compounds are itinerant antiferromagnetic (AFM) semimetals.
     SC in these compounds emerges upon suppression of the stripe-type AFM phase by application of pressure and/or carrier doping, where the AFM spin fluctuations are still strong. 
    Therefore, it is believed that stripe-type AFM spin fluctuations play an important role in driving the SC in the iron-based superconductors, although orbital fluctuations are also pointed out  to be important. \cite{Kim2013}
   Recently ferromagnetic (FM) correlations were revealed to play an important role in the iron-based superconductors. \cite{Johnston2010,Nakai2008,PaulPRB, PaulPRL}

      The relationship between crystal structure and SC is also an important issue in the Fe-based superconductors.
     Many $``$122$"$ compounds are reported to form in the body-centered tetragonal (${\cal T}$) ThCr$_2$Si$_2$-type structure  (space group $I4/mmm$) at room temperature and to exhibit a structural phase transition from the ${\cal T}$ structure to an orthorhombic (${\cal O}$) one on cooling. \cite{Johnston2010,Canfield2010,Stewart2011}
    CaFe$_2$As$_2$ is one such compound exhibiting an AFM ordering of the Fe moments at $T_{\rm N}$ = 170 K  with a concomitant structural phase transition from a high-temperature (HT) ${\cal T}$ to a low-temperature (LT) ${\cal O}$ phase.\cite{Ni2008, Goldman2008, Canfield2009} 
    However, the collapsed-tetragonal (c${\cal T}$) phase can be stabilized by modest pressure application. \cite {Canfield2009,Torilachvili2008,Lee,Yu}
    The c${\cal T}$ phase in CaFe$_2$As$_2$ is characterized by a $\sim$10 $\%$  reduction in the tetragonal $c$ lattice constant, from the value in the HT ${\cal T}$ phase, along with the absence of the LT ${\cal O}$ AFM ordered phase. \cite{Kreyssig2008,Goldman2009_2, Ran2011}
   The c${\cal T}$ phase in CaFe$_2$As$_2$ can also be induced by rare-earth (Nd, Pr) substitution at the Ca site or by changing the heat treatment conditions that control strains inside a crystal grown out of excess FeAs due to the formation of nanoscale precipitates. \cite{Ran2011, Ran2012, Saha2012}
      No SC was observed in the c${\cal T}$ phase of CaFe$_2$As$_2$ down to 100 mK. \cite{CaFe2As2-unpublish}

     The complete substitution of the Fe atoms in CaFe$_2$As$_2$ by the 4$d$ element Pd which has higher number of outer-shell $d$ electrons than that of Fe is found to induce a c${\cal T}$ structure. \cite{CaPd2As2} 
      From detailed magnetic susceptibility $\chi$, magnetization, specific heat, in-plane resistivity, and in-plane magnetic penetration depth measurements,  CaPd$_2$As$_2$ is revealed to be a conventional type-II nodeless $s$-wave superconductor with $T_{\rm c}$ = 1.27 K. \cite{CaPd2As2}   
    Therefore it is interesting to investigate the magnetic and electronic properties of the CaPd$_2$As$_2$ superconductor, especially focusing on magnetic correlations in the normal state. 

     Nuclear magnetic resonance (NMR) is known to be a microscopic probe suitable for investigating static spin susceptibility and  low-energy spin excitations for pnictide superconductors. \cite{Johnston2010,Ishida2009,Ma2013}
     It is known that the temperature $T$ dependence of the nuclear spin-lattice relaxation rate (1/$T_1$)  reflects the wave vector $q$-summed dynamical susceptibility. 
     On the other hand, NMR spectrum measurements, in particular the Knight shift $K$, give us information on static magnetic susceptibility $\chi$. 
    Thus from the $T$ dependence of 1/$T_1T$ and $K$, one can obtain valuable insights about magnetic fluctuations in materials.  
     Furthermore, 1/$T_1$ measurements in the SC state provide important information in understanding the gap structure in SCs.        

    In this paper we report $^{75}$As NMR and nuclear quadrupole resonance (NQR) measurements to examine the magnetic fluctuations in CaPd$_2$As$_2$. 
    From the $T$ dependence of 1/$T_1$ and the $K$'s, the absence of AFM spin correlations is clearly evidenced in the normal state, which is in contrast to Fe-based superconductors.    
   In the SC state,  1/$T_1$ shows a clear Hebel-Slichter (HS) peak just below $T_{\rm c}$ and decreases exponentially at lower temperatures, confirming an $s$-wave SC. 
    In addition, we explain the clear suppression of the HS peak with applied magnetic field $H$ in terms of the Doppler shift effect, also known as Volovik effect, \cite{Volovik}  in the $s$-wave SC.  
     To our knowledge, this is the first observation of the effect in a full-gap $s$-wave SC, although the Volovik effect in gapless nodal superconductors has been observed.\cite{specific_heat,  NMR_volovik0}

 
    The single crystals of  CaPd$_2$As$_2$ for the NMR measurements were grown using PdAs self-flux as reported in detail elsewhere. \cite{CaPd2As2}     
   NMR and NQR measurements were carried out on $^{75}$As  (\textit{I} = 3/2, $\gamma/2\pi$ = 7.2919 MHz/T, $Q$ =  0.29 barns)  by using a homemade, phase-coherent, spin-echo pulse spectrometer. 
   The $^{75}$As-NMR spectra were obtained by sweeping $H$ at a fixed frequency $f$ = 53 MHz, while $^{75}$As-NQR spectra in zero field were measured in steps of frequency by measuring the intensity of the Hahn spin echo.  
       In the SC state, since a low upper critical field $H_{c2}$(0) of 1.57 kOe makes NMR investigations in the SC state very difficult,\cite{CaPd2As2} we have carried out NQR measurements in $H=0$.   
  Below $T_{\rm c}$ we only performed NQR measurements using powdered samples as intensities of NMR/NQR signals with the single crystals decrease drastically making the measurements difficult in the SC state.
   The $^{75}$As 1/$T_{\rm 1}$ was measured with a saturation recovery method. \cite{T1}

\begin{figure}[tb]
\includegraphics[width=8.5cm]{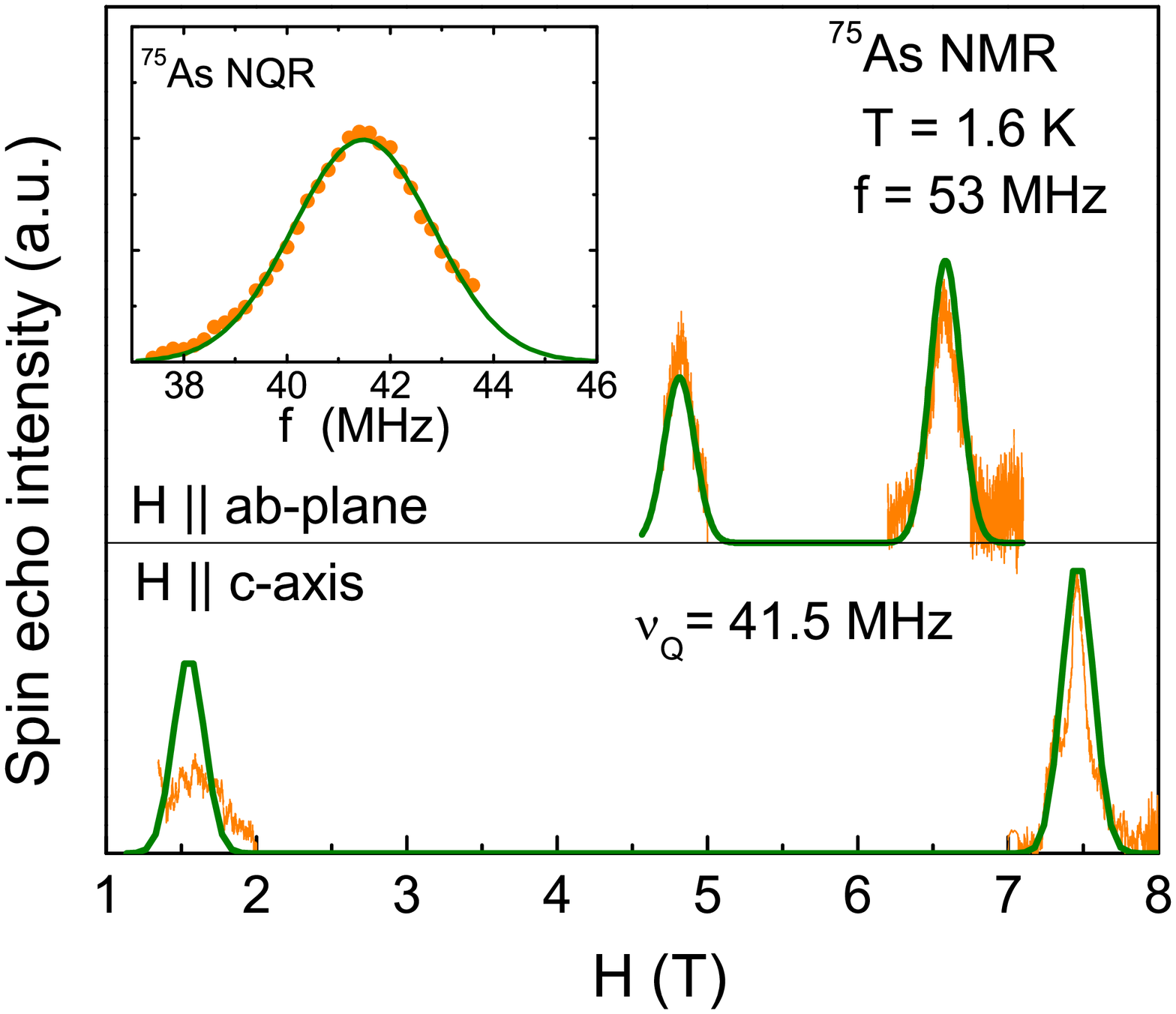} 
\caption{(Color online) (a) Field-swept $^{75}$As-NMR spectra of a CaPd$_2$As$_2$ crystal at $f$ = 53 MHz and $T$ = 1.6 K  for $H$ $\parallel$ $c$ axis (bottom) and $H$ $\parallel$ $ab$ plane (top).   
    The orange and green lines are the  observed and simulated spectra, respectively.  
     Inset: $^{75}$As NQR spectrum at $T$ = 1.6 K and $H$ = 0 T.
 }
\label{fig:As-spectrum}
\end{figure}

    Figure\ \ref{fig:As-spectrum} shows typical field-swept $^{75}$As-NMR spectra of a CaPd$_2$As$_2$ crystal   at  $T$ = 1.6 K   for two magnetic field directions, $H$ $\parallel$ $c$ axis  and $H$ $\parallel$ $ab$ plane. 
   The typical spectrum for a nucleus with spin $I=3/2$ with Zeeman and quadrupolar interactions can be described by a nuclear spin Hamiltonian ${\cal{H}}=-\gamma\hbar(1+K)HI_z+\tfrac{h\nu_Q}{6}[3I_z^2-I(I+1)]$,
where $h$ is Planck's constant, and $\hbar = h/2\pi$. 
   The nuclear  quadrupole frequency for $I=3/2$ nuclei is given by $\nu_{\rm Q} = e^2QV_{\rm ZZ}/2h$, where $Q$ is the nuclear quadrupole moment and $V_{\rm ZZ}$ is the electric field gradient at the As site.
    For $I=3/2$ nuclei, this Hamiltonian produces a spectrum with a sharp central transition line flanked by one satellite peak on both sides. 
  The observed $^{75}$As NMR spectra are well reproduced by simulated spectra (green lines in Fig.\ \ref{fig:As-spectrum}) from the above simple Hamiltonian with  $\nu_{\rm{Q}}$ = 41.5 MHz, and Knight shifts $K_{ab}= (0.32\pm 0.05)\%$  for $H$ $\parallel$ $ab$ plane and $K_c =  (0.37\pm0.05)\%$ for $H$ $\parallel$ $c$ axis.   
   These NMR parameters do not depend on temperature. 
   Here the expected satellite lines above 8.5 T are not measured due to the limited maximum $H$ of our SC magnet.
   The $T$-independent $K$s are consistent with the nearly $T$-independent $\chi$s for both $H$ directions reported in Ref. \onlinecite{CaPd2As2}.
   The value of  $\nu_{\rm Q}$ = 41.5 MHz is confirmed by the observation of an NQR spectrum at $H=0$ and $T$ = 1.6 K shown in the inset of Fig.\ \ref{fig:As-spectrum}.
   A similar value of $\nu_{\rm Q}$  = 41.1 MHz was reported in the  c${\cal T}$ phase in CaFe$_2$As$_2$.\cite{Furukawa}
   Comparable $\nu_{\rm Q}$ values of $\nu_{\rm Q}$  $\sim$ 35.8 and 41.5 MHz  were also reported in the c${\cal T}$ phase of (Ca$_{1-x}$Pr$_x$)Fe$_2$As$_2$ for $x$ = 0.075 and 0.15, respectively. \cite{Ma2013_2}

\begin{figure}[tb]
\includegraphics[width=8.5cm]{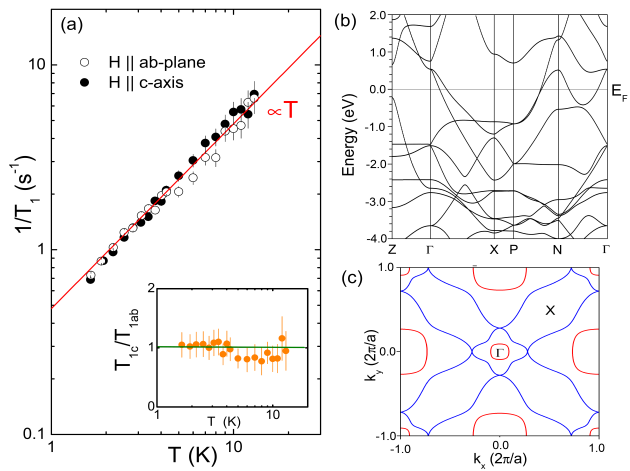} 
\caption{(Color online) 
    (a) $T$ dependence of  1/$T_1$  for CaPd$_2$As$_2$ for both $H$ $\parallel$ $c$ axis and $H$ $\parallel$ $ab$ plane.  
   The straight line shows the Korringa dependence 1/$T_1$ $\propto$ $T$. 
Inset: $T$ dependence of the ratio $R$ $\equiv$ $T_{\rm 1,c}$/$T_{\rm 1,ab}$.
    (b) Band dispersion of CaPd$_2$As$_2$. 
    (c) Cross-section of the Fermi surface at $k_z=0$. Blue and red curves correspond to two different bands. }
\label{fig:T1-NMR}
\end{figure}

    Figure\ \ref{fig:T1-NMR} shows the $T$ dependence of 1/$T_1$ for  $H$ $\parallel$ $c$ axis  and $H$ $\parallel$ $ab$ plane.
    As shown by the red solid line, 1/$T_1$ varies in proportion to $T$ with no anisotropy in the normal state, following the Korringa relation 1/$T_1T$ = constant as expected for a Fermi liquid. 
     No enhancements in  1/$T_1T$ and $T$-independent $K$s are clear evidence of no strong electron correlations in the c${\cal T}$ phase in CaPd$_2$As$_2$.
     A similar quenching of magnetic correlations has been reported in the c${\cal T}$ phase of CaFe$_2$As$_2$.\cite{Furukawa, Kawasaki2010}

    In order to evaluate more quantitatively the strength of the electron correlations, it is useful to estimate the quantity $T_1TK_{\rm s}^2$ where $K_ {\rm s}$  is the spin part of the Knight shift. \cite{Moriya1963,Narath1968,Li2010} 
 The so-called Korringa ratio 
 $S = 4\pi k_{\rm B} T_1TK_{\rm s}^2\gamma_{\rm n}^2/\hbar\gamma_{\rm e}^2$
is unity for uncorrelated metals. 
 Here  $\gamma_{\rm e}$ and $\gamma_{\rm n}$ are the electron and nuclear gyromagnetic ratios, respectively. 
   For strongly AFM-correlated metals, $S$ $\ll$ 1. 
   For FM-correlated metals, on the other hand, $S$ $\gg$ 1.
   Assuming $S$ = 1 for uncorrelated metals, $K_{\rm s}$ is estimated to be 0.2 $\%$ using the 1/$T_1T$ value in the normal state. 
   The $K_{\rm s}$ would be consistent with the observed total $K$ = $K_{\rm s}$ +$K_{\rm 0}$ if we subtract the typical value of the orbital part of $K_0$ = 0.1 $\sim$ 0.2 $\%$ in iron pnictides. \cite{PaulPRL}

   Furthermore, the absence of spin correlations can be seen by looking at the ratio $R$ $\equiv$ $T_{1,c}$/$T_{1, ab}$. 
It is known that this ratio depends on magnetic correlations in Fe pnictides. \cite{KitagawaSrFe2As2,SKitagawaAF-Fluctuation,FukazawaKBaFe2As2, Furukawa}
  In most of Fe pnictide SCs, $R$ is greater than unity corresponding to the stripe-type AFM fluctuations.\cite{KitagawaSrFe2As2,SKitagawaAF-Fluctuation,FukazawaKBaFe2As2, Furukawa}
    On the other hand, $R$ = 1 is expected for uncorrelated systems, as has been actually observed in the c${\cal T}$ phase of the non-SC CaFe$_2$As$_2$. \cite{Furukawa}
    As plotted in the inset of Fig.\ \ref{fig:T1-NMR}(a), the ratio is almost constant with $R$ $\sim$ 1.
Again this is consistent with the absence of spin correlations in CaPd$_2$As$_2$, in contrast to the case of Fe-based superconductors where the stripe-type AFM spin correlations are believed to play an important role in the appearance of unconventional superconductivity. 
    Since the stripe-type AFM spin correlations originate from the interband correlations due to the multi-band nature of the Fermi surface,  the absence of the stripe-type AFM spin correlations in CaPd$_2$As$_2$ indicates a drastic difference  in Fermi surface topology. 

   To understand the band structure of  CaPd$_2$As$_2$, we performed electronic structure calculations\cite{Yongbin}  using the full-potential linearized augmented plane wave  method \cite{Blaha2001} with a generalized gradient approximation.\cite{Perdew1996}
    The calculated band dispersion is shown in Fig.\ \ref{fig:T1-NMR}(b), which is in good agreement with the previous report. \cite{Nekrasov2013}
    We found two hole pockets around the ${\it \Gamma}$ point and no electron pocket at the  ${\it X}$ point along the [1,1,0] direction [see Fig.\ \ref{fig:T1-NMR}(c)],   making no interband correlations possible.   
   This is consistent with the absence of the AFM spin correlations revealed by the 1/$T_1$ measurements.
    It is interesting to point out that a similar quenching of the AFM spin correlations has been observed in the non-SC c${\cal T}$  phase of CaFe$_2$As$_2$ where the hole pockets around the ${\it \Gamma}$ point sink below the Fermi energy.\cite{Dhaka2014}

\begin{figure}[tb]
 \includegraphics[width=8.5cm]{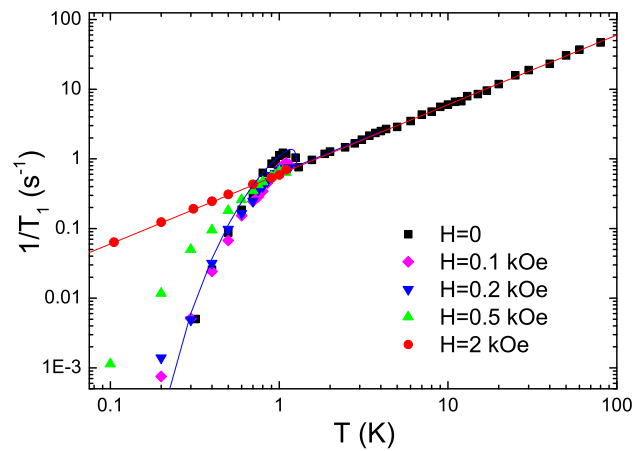} 
  \caption{(Color online) $T$ and $H$ dependence of the $^{75}$As-NQR 1/$T_1$. 
   The red straight line  represents a 1/$T_1T$ = constant relation. 
    The blue solid curve below $T_{\rm c}$ is calculated based on the BCS theory.  }
  \label{fig:T1-NQR}
  \end{figure}

  Now we discuss the SC properties below $T_{\rm c}$ = 1.27~K. 
      As shown in Fig.\ \ref{fig:T1-NQR}, under zero field, 1/$T_1$ shows a clear coherence peak (Hebel-Slichter, HS, peak) just below $T_{\rm c}$ and decreases exponentially at low temperatures, which are characteristics of conventional BCS superconductors. 
   The blue solid curve in Fig.\ \ref{fig:T1-NQR} is a calculation using the BCS theory. 
  Here, the relaxation rate 1/$T_{\rm 1s}$  in the SC state normalized by $1/T_{\rm 1n}$ in the normal state is expressed as \cite{Hebel1959}
 \begin{eqnarray}
\frac{T_{\rm 1n}}{T_{\rm 1s}} \propto \int_{0}^{\infty}[{N_{\rm s}(E)}^2+{M_{\rm s}(E)}^2]f(E)[1-f(E)]dE
 \end{eqnarray}
where $M_{\rm s}(E)$ = $N_0\Delta$/$\sqrt{E^2-{\Delta}^2}$ is the anomalous density of states (DOS) due to the coherence factor,  $N_{\rm s}(E)$ = $N_{\rm 0}E$/$\sqrt{E^2-{\Delta}^2}$ is the DOS in the superconducting state, $\Delta$ is the energy gap, $N_0$ is the DOS in the normal state, and $f(E)$ is the Fermi distribution function. 
   We convolute  $M_{\rm s}(E)$ and $N_{\rm s}(E)$ with a broadening function assuming a triangle shape with a width 2$\delta$ and a height 1/$\delta$. 
  Using  2$\Delta(0)/k_{\rm B}T_{\rm c}$ = 3.16 from Ref.~\onlinecite{CaPd2As2} and  $r=\Delta(0)/\delta=5$, 
the experimental data were reproduced reasonably well as shown by the blue curve in Fig.\ \ref{fig:T1-NQR}. 
    The value of 2$\Delta(0)/k_{\rm B}T_{\rm c}$ is slightly smaller than that of the BCS weak-coupling limit 2$\Delta(0)/k_{\rm B}T_{\rm c}$ = 3.53, which originates from the SC gap anisotropy in CaPd$_2$As$_2$ reported previously.\cite{CaPd2As2}

\begin{figure}[tb]
\includegraphics[width=7.5cm]{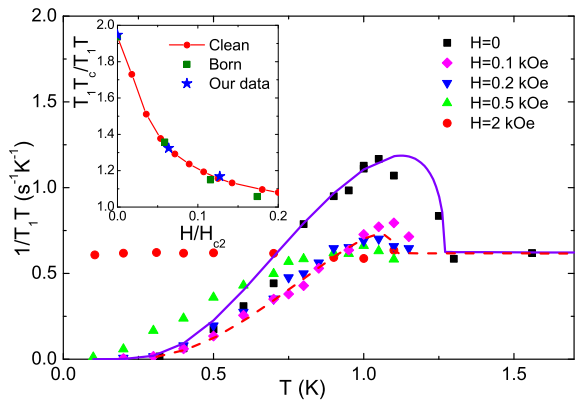} 
\caption{(Color online) $T$ and $H$ dependence of the $^{75}$As 1/$T_1T$ measured by NQR. 
  The straight line above $T_{\rm c}$  represents a 1/$T_1T$ = constant relation. 
  The solid curve below $T_{\rm c}$ is a calculation assuming the BCS model. 
   The dotted curve is a calculation without the coherence term $M_{\rm s}(E)$.   
   Inset: Height of the HS peak as a function of $H/ H_{\rm c2}$.  
   Star points represent our experimental data. 
    The solid line and square points are results calculated by Tanaka ${\it et~al.}$, based on Eilenberger theory. \cite{Tanaka}
     The solid line is for the clean limit and square points are for the Born limit. See Ref. \onlinecite{Tanaka} for details.}
\label{fig:T1-SC}
\end{figure}

    In the presence of low $H$ of 0.1 kOe and 0.2 kOe,  larger than the lower critical field $H_{\rm c1}$(0) of 14 Oe,\cite{CaPd2As2} 1/$T_1T$ does not show much change at low $T$, but we observe a clear suppression of the HS peak with increasing $H$.
   To see the suppression more clearly, we show the low-$T$ part of the 1/$T_1T$ data in Fig.\ \ref{fig:T1-SC}.  
    The  1/$T_1T$ at $H$ = 0.5 kOe is greater than the intrinsic BCS rate at low $T$, which is due to spin diffusion effects where the nuclear spin polarization diffuses to the vortices (normal cores)  in the mixed state. \cite{Silbernagel} 
    When a magnetic field of 2 kOe (higher than the $H_{\rm c2} = 1.57$ kOe) was applied, the SC was completely suppressed and then Fermi liquid behavior 1/$T_1T$ = constant is observed.

In the inset of Fig.\ \ref{fig:T1-SC}, the $H$ dependence of the HS peak height is presented. The peak height decreases rapidly with increasing $H$.
The suppression of the HS peak by increasing $H$ has also been observed in type-II superconductors V$_3$Sn, LaFe$_4$P$_{12}$ and fulleride superconductors. \cite{Okubo, Masuda, Nakai2005, Pennington}
   Qualitatively speaking, the suppression of the HS peak in 1/$T_1$ is due to smearing of the gap-edge shape of the DOS in the SC state: the greater the broadening, the more suppressed is the HS peak.    
   Several mechanisms have been proposed so far: (1)  broadening of the gap edge due to a finite lifetime of Cooper pairs \cite{Pennington, Cryot} and (2) Zeeman energy effects on the gap edge for quasiparticle states. \cite{Pennington}

   Cyrot ${\it et~al}$., have calculated the lifetime effects on 1/$T_1$ in the mixed state and obtained the formula 
$\frac{d(T_{1n}/T_{1s})}{dT}\vert _{T=T_c(H)}= \frac{ec}{\sigma}\frac{dH_{c2}/dT}{\beta_A(2\kappa^2-1)}\frac{1}{k_BT_c} g(t).$ \cite{Cryot}
Here $\sigma$ is the normal-state electrical conductivity, $\beta_A$ is a coefficient of order 1, $\kappa$ is the Landau-Ginzberg parameter,  $t=T_{\rm c} (H)/T_{\rm c}$, $T_{\rm c}(H$) is the suppressed value of $T_{\rm c}$ in the applied field $ H$, and $g(t)$ is computed numerically by Cyrot.\cite{Cryot} 
   Although the main features of the prediction have been verified in conventional superconductors for $H$ near $H_{c2}$ by Masuda and Okubo, \cite{Okubo, Masuda} they found that the formula does not explain their data at low $H$. 
    We also found that for the case of CaPd$_2$As$_2$, estimates of the quantity $\frac{d(T_{1n}/T_{1s})}{dT}$  are typically one order of magnitude larger than the observed experimental value.
   This indicates that the broadening of the gap edge due to the lifetime effect is not enough to explain the suppression in our case.
   As for the second Zeeman energy shift effects, in our case, the Zeeman energy of the applied field of 200 Oe is just about 1\% compared to the SC energy gap. 
   Therefore, these effects cannot be responsible for the strong suppression of the HS peak observed in CaPd$_2$As$_2$.

   Thus one needs to introduce another effect. 
   Here we consider the Doppler shift effect of qusiparticle excitations, also known as Volovik effect.\cite{Volovik}
     The idea of the effect is simple:~the quasiparticle energy spectrum is $``$Doppler$"$  shifted in the mixed state due to a supercurrent flow $v_{\rm s}$ around the vortices as $E(p) \to E(p) + {\vec v_{\rm s}}\cdot {\vec p} $.
   This leads to a broadening of the gap edge structure, resulting in the suppression of the HS peak.  
    Although  the energy shift due to the Volovik effect is typically very small in comparison with the SC gap energy, the Volovik effect has been observed in many gapless nodal SCs. \cite{specific_heat,  NMR_volovik0,NMR_volovik1} 
    This is due to the fact that the Doppler shift produces a finite DOS at the Fermi energy ($E_{\rm F}$)  because of the $k$-linear dependence of the DOS in gapless nodal SCs. 
    On the other hand, the effect does not produce a finite DOS at $E_{\rm F}$  in full-gap $s$-wave SCs.  
    Therefore, the Volovik effect has not yet been discussed in full-gap $s$-wave SCs, although a possible observation of the effect has been proposed in a more complex $s$-wave SC having multiple gaps with different magnitudes.\cite{NMR_volovik2}
    However, even for the full-gap SC, the Volovik effect can be detected by looking at how the gap edge of the DOS is smeared by the application of magnetic field. 
   The maximum Doppler shift energy is estimated to be $\sim$ 0.5 meV in CaPd$_2$As$_2$, which is comparable to the SC gap energy $\Delta(0)$ $\sim$ 0.2 meV. 
    Therefore, we may expect that the HS peak is suppressed strongly due to the Volovik effect. 
    In fact, the 1/$T_1$ data at low $H$ can be reproduced only if we exclude the coherence term $M_{\rm s}(E)$ in the calculation (see the dotted curve in  Fig.\ \ref{fig:T1-SC}) assuming a very strong suppression of the HS peak.    
   Since such strong suppression of the HS peak cannot be attributed to the two well-known effects described above, we conclude that the suppression is mainly due to the Volovik effect, providing the experimental observation of the effect in a full-gap $s$-wave SC.     
   To our knowledge, this is the first explanation of the suppression of HS peak in $s$-wave SC in 1/$T_1$ versus $H$ in terms of the Volovik effect.

   Quite recently Tanaka and coworkers have quantitatively studied  the effects of vortices on $H$ and $T$ dependences of $1/T_1$ in the mixed state based on Eilenberger theory. \cite{Tanaka}
   Square points in the inset of  Fig.\ \ref{fig:T1-SC}  show the calculated $H$ dependence of the peak height in the presence of impurity scattering in the Born limit. \cite{Tanaka} 
  They also calculated the $H$ dependence  of the peak height in the clean limit which shows similar behavior, as shown by the solid line. \cite{Tanaka}
  In CaPd$_2$As$_2$, the Ginzburg-Landau coherence length at $T$ = 0, $\xi (0)$ = 45.8 nm, is much larger than the mean-free path $l$ = 1.52 nm, \cite{CaPd2As2} indicating that CaPd$_2$As$_2$ is in the dirty limit. 
    Our experimental data on CaPd$_2$As$_2$ in Fig.\ \ref{fig:T1-SC} show excellent agreement with the theoretical values predicted by Eilenberger theory. 

    In summary, we have performed $^{75}$As NMR and NQR measurements on the c${\cal T}$-phase CaPd$_2$As$_2$ superconductor. 
    Similar to the c${\cal T}$ phase of CaFe$_2$As$_2$, the absence of spin correlations has been shown in CaPd$_2$As$_2$ by the Korringa relation 1/$T_1T$ = constant and Korringa ratio $S$ close to 1. 
   The nuclear spin-lattice relaxation rate 1/$T_1$ shows a Hebel-Slichter peak below $T_{\rm c}$ and decreases exponentially at low temperatures, indicating that CaPd$_2$As$_2$ is a full-gap $s$-wave superconductor. 
    We attribute the suppression of the HS peak under magnetic field in this full-gap SC to the supercurrent Doppler shift (Volovik)  effect. 
    Although Doppler effects have been discussed in gapless nodal SCs, we conclude that the effect can play an important role in the nature of SC in the mixed state even for full-gap $s$-wave superconductors.

  We thank Masanori Ichioka  for helpful discussions. 
The research was supported by the U.S. Department of Energy, Office of Basic Energy Sciences, Division of Materials Sciences and Engineering. Ames Laboratory is operated for the U.S. Department of Energy by Iowa State University under Contract No.~DE-AC02-07CH11358.


\begin{thebibliography}{10}
\bibitem{T1}  $1/T_1$ at each $T$ was determined by fitting the nuclear magnetization $M$ versus time $t$  using the exponential functions $1-M(t)/M(\infty) = 0.1 e^ {-t/T_{1}} +0.9e^ {-6t/T_{1}}$ for $^{75}$As NMR,  and  $1-M(t)/M(\infty) = e^ {-3t/T_{1}}$ for $^{75}$As NQR,  where $M(t)$ and $M(\infty)$ are the nuclear magnetization at time $t$ after the saturation and the equilibrium nuclear magnetization at $t$ $\rightarrow$ $\infty$, respectively. 
    In the paramagnetic state, the nuclear magnetization recovery curves were well fitted by the function within our experimental uncertainty. 
     On the other hand, below $T_{\rm c}$ = 1.27 K, we observed a slight deviation from the function under magnetic fields and fitted with two $T_1$ components  and took the long component as $T_1$.
\end{thebibliography}
\end{document}